\documentstyle[12pt]{article}

\newcommand{\bs}{\bigskip}

\newcommand{\ba}{\begin{array}}
\newcommand{\ea}{\end{array}}
\newcommand{\eps}{\epsilon}

\newcommand{\ra}{\rightarrow}

\newcommand{\be}{\begin{equation}}
\newcommand{\ee}{\end{equation}}
\newcommand{\bea}{\begin{eqnarray}}
\newcommand{\eea}{\end{eqnarray}}

\begin{document}

\centerline{\Large \bf Fivebrane Lagrangian with}

\bigskip

\centerline{\Large \bf  Loop Corrections in  Field-Theory Limit
\footnote{Research supported by the
ISF Grant MOY000 and by RFFI Grant 95-02-05822a} }

\vspace{1cm}

\centerline{\bf N.A.Saulina $^{a,b,}$\footnote{e-mail:
saulina@vxitep.itep.ru},
M.V.Terentiev $^{a,}$\footnote{e-mail: terent@vxitep.itep.ru},
K.N.Zyablyuk $^{a,}$\footnote{e-mail: zyablyuk@vxitep.itep.ru}}

\bigskip

\centerline{$^a \ \ Institute \ \ of \ \ Theoretical \ \
 and \ \  Experimental \ \ Physics $  }
\centerline{$^b \ \  Moscow \ \ Physical \ \ Technical \ \ Institute $}

\bigskip

\begin{abstract}
Equations of motion and the lagrangian are derived explicitely
for  Dual D=10, N=1
Supergravity considered as a field theory limit of a Fivebrane.
It is used the mass-shell solution of Heterotic String Bianchi Identites
obtained in the  2-dimensional
$\sigma$-model two-loop approximation and in the  tree-level Heterotic
String approximation.
As a result the Dual Supergravity lagrangian is  derived in the
one-loop Five-Brane approximation and in the lowest
6-dimensional $\sigma$-model approximaton.

\end{abstract}

\section{Introduction}

There are two kinds of parameters, characterising the  heterotic superstring
field-theory limit. The {\it first parameter}  is the string-tension
$\alpha'$ which enters into the $\sigma$-model lagrangian, corresponding
to the superstring tree-level:

$$ L_{HS} = -{1\over 2\pi \alpha'}\int d^2  \xi \,
\left({1\over 2}{\sqrt{-\gamma}}\gamma^{ij}\partial_i  X^m \partial_j X^n
g_{mn}^{HS}(X)+ \right. $$
\be
\label{LHS}
\left.+{1\over 2}\eps^{ij}\partial_i  X^m \partial_j X^n
B_{mn}(X)+ \ldots \right)
\ee
Here  $\alpha'$  is the $ \sigma  $-model loop expansion parameter,
$R^{(2)}$-is the 2-dimensional curvature.

The Type I supergravity lagrangian that follows from (\ref{LHS}) in the
field-theory limit takes the form:

\be
\label{LSG}
L_{SG} ={{\sqrt {-g^{HS}}}\over 2k^2}\exp(-2\varphi)(R +
8(\partial_n  \varphi)^2 +{1\over 12}K_{mnp}^2 + \ldots)
\ee
Here $k^2$ is the Newton gravitational constant, $\varphi$ is the dilatonic
field, $R$ is the curvature scalar.  All the entries in
(\ref{LSG}) are calculated in terms of string metric: $g_{mn}^{HS} =
exp(\varphi/2)g_{mn}$, where $g_{mn}$ is a canonical metric tensor;
$K$ is the 3-form axionic gauge-field, corresponding to the
2-form potential $B$ in (\ref{LHS}):

\be
\label{H}
K= dB + 2\alpha' (\Omega_G -\Omega_L)
\ee
where $\Omega_G$ and $\Omega_L$ are Chern-Symons (CS) 3-form terms,
corresponding to the gauge-group ($G$) and Lorentz $O(1,9)$ group ($L$).

The {\it second parameter} of the Heterotic String theory is
$\sim  k^2 /\alpha'$.
It is the string loop expansion parameter. The general form of the
supergravity  lagrangian in the superstring field-theory limit
 takes the form in terms of string metric \cite{DL1}:

$$
L_{SG} ={\sqrt {-g^{HS}}\over 2k^2}\exp(-2\varphi)
\sum_{l=1,2 \ldots} \sum_{L=0,1,\ldots}L_{L,l-1}
$$
\be
\label{LSGE}
L_{L,l-1}=a_{L,l-1} (\alpha')^{l-1}
 \left({2\,k^2 \over \exp(-2\varphi)(2\pi)^5\alpha'}\right)^L R^{l+3L}
+\ldots
\ee

Here $L$ -is a number of string loops, $l$ is a number of
$\sigma$-model loops. The value $L=0$ corresponds to the
(classical)  heterotic string  tree-level.

We keep only terms with curvature in (\ref{LSGE}). Terms with the gauge-field
appear  in a series of similar structure. To obtain the generic
structure of all the terms one must keep in mind the
dimensions (in mass units): $dim(\alpha') =-2 $, $dim(k^2)=-8 $.

Note that terms, which are presented in (\ref{LSG})
correspond to all the bosonic contributions at
$l=1 $ and $L=0 $ level. The CS-terms in (\ref{H}) take into
account the (anomaly cancelling)  $l=2$ contribution \cite{GS}.

Now we turn to the fivebrane.  In the $\sigma$-model limit it
is described by the lagrangian:

$$ L_{FB} = -{1\over (2\pi)^3 \beta'}\int d^6  \xi \,
\left({1\over 2}{\sqrt{ -\gamma}}\gamma^{ij}\partial_i  X^m \partial_j X^n
g_{mn}^{FB}(X)+\right. $$
\be
\label{LFB}
\left. +{1\over 6}\eps^{j_1\ldots j_6}\partial_{j_1}  X^{m_1}
\ldots  \partial_{j_6} X^{m_6}
C_{m_1 \ldots m_6}(X) + \ldots\right)
\ee
Here $\beta'$ is the fivebrane $\sigma$-model loop expansion parameter.
It is expected, that the  dual supergravity lagrangian that follows from
(\ref{LFB}) is dual to (\ref{LSG}) and takes the form:

\be
\label{LDSG}
L_{DSG} ={{\sqrt {-g^{FB}}}\over 2k^2}\exp(2\varphi /3)(R +
{1\over 2\cdot 7!}M_{n_1 \ldots n_7}^2 + \ldots)
\ee

Here $M$ is the 7-form gauge field, corresponding to the
6-form potentials $ C$ in (\ref{LFB}) and dual to $K$. All the entities in
(\ref{LDSG}) are calculated in terms
of  the fivebrane metric: $ g^{FB}_{mn}=\exp(-\varphi/6)g_{mn}$.
The  $M$-field  is defined by:

\be
\label{M}
M= dC + \beta'\, X_7
\ee
Here $X_7$ is the Chern-Simons 7-form, containing curvatures and
connections of $G$ and $O(1.9)$-groups \cite{SS}, \cite{GN2}.

The general structure of dual supergravity lagrangian takes the
form  \cite{DL1}:
$$
L_{DSG} ={{\sqrt {-g^{FB}}}\over 2k^2}\exp(2\varphi /3)
 \sum_{l'=1,2 \ldots} \sum_{L'=0,1,\ldots}L'_{L',l'-1}
$$
\be
\label{LDSGE}
L'_{L',l'-1}={a'}_{L',l'-1} (\beta')^{l'-1}
({2\,k^2 \over \exp(2\varphi /3)(2\pi)^5\beta'})^{L'}
R^{L'+3l'-2}+\ldots
\ee
Here $2\,k^2/ \exp(2\varphi/3)(2\pi)^5\beta'$ is the fivebrane
loop expansion
parameter, but  $l'$ and $L'$ are interpreted as  numbers of 6-dimensional
$\sigma$-model
loops and fivebrane  loops respectively. The value $L'=0$
corresponds to the (classical) fivebrane tree-level.

In (\ref{LDSG})  all the terms are kept,
that correspond to the
$l'=1$,  $L'=0$ level. But the CS-terms in (\ref{M}) take into account
$l'=2$, $L'=0$  contribution.

The discovery of the heterotic string - fivebrane duality \cite{S}, \cite{D},
\cite{DL2} leads presumably to the equality of two series (\ref{LSGE})
and (\ref{LDSGE}).
There is the  relation between expansion parameters in these
theories \cite{DL1}:
\be
\label{PAR}
\beta' = {2k^2 \over (2\pi^5)\alpha' }
\ee
This relation makes it possible to establish term-by-term
correspondence between
(\ref{LSGE}) and (\ref{LDSGE}). That leads to the statement:

\be
\label{ll'LL'}
a_{Ll}= {a'}_{L'l'}, \ \ \ L' = l-1, \ \ L = l'-1
\ee
It means that $\sigma$-model loop expansion in the heterotic string
case  reproduces the
loop expansion in the fivebrane theory;
the loop expansion in the heterotic string case reproduces the
$\sigma$-model expansion in the fivebrane theory.

Then the String/Fivebrane symmetric form of effective lagrangian
follows from (\ref{PAR}) and (\ref{ll'LL'}) (in terms of canonic metric):

\be
\label{UN}
L_{S/FB} ={{\sqrt {-g}}\over 2k^2}
\sum_{L=0}^\infty \sum_{L'=0}^\infty \, a_{LL'}
(\alpha')^{L'}(\beta')^L \exp(\varphi(L-L')/2)R^{3L+L'+1}+ \ldots
\ee

It is not simple to   check independently
this beautiful statement, because  fivebrane theory has not
quantized. Moreover,  even the consisent supersymmetric fivebrane
lagrangian has not yet constructed.

We hope, that the supersymmetry might
present additional insight to the problem.
We calculate in the present paper, accepting the picture described above,
the supersymmetric one-loop corrections to the
dual supergravity lagrangian from  the second-loop $\sigma$-model
corrections in the Type I D=10, N=1 supergravity,
considered as the field-theory limit  of heterotic superstring.

We get: 1) the supersymmetry transformations, 2) the explicitely
supersymmetric equations of motion, and  3) the supersymmetric  lagrangian
for the fivebrane at the $l'=1$, $L'=1$ level in the field-theory limit
(that corresponds to the dual supergravity with specific loop corrections).
\footnote{As for lagrangian, we are able to present  all the
bosonic terms. The fermionic terms are too complicated to be presented
explicitely. It is a technical problem because one can construct any
desired term using the described procedure}

It is interesting  in the framework of described approach
to obtain  all the one-loop supersymmetric
heterotic string corrections from the calculation of
$\sim \beta'$ terms  in the dual supergravity. It is a problem for
the future.

The dual supergravity lagrangian obtained in the present paper
corresponds to the supersymmetrised version of $\sim \alpha' $
 anomaly cancelling Green-Schwarz (GS)  corrections \cite{GS}  to the
simple (lowest $\alpha'$-order) D=10, N=1  supergravity considered in
\cite{CM}. (That corresponds to
the $l=2$, $L=0 $ level  of heterotic string expansion).
The  problem with  the standard supergravity  is
that $\sim \alpha'$ corrections
can be made  supersymmetric only in the same $\sim \alpha' $ order.
For complete
supersymmetrisation one must take into account the infinite number of terms
$\sim \alpha'^n, \ \ (n = 1,2,\ldots), $ containing the axionic field and
dictated by supersymmetry.
Situation is different in the dual supergravity. If the same
corrections are expressed in terms of fivebrane variables - the result
becomes exactly supersymmetric in the order $ \sim \alpha' $ , i.e. the
infinite series in $\alpha'$  is transformed to the finite number of
terms in the case of dual supergravity.
That means at least that the relation (\ref{ll'LL'})
can not be applied directly
to  terms containing  axionic field.

The supersymmetric completion of the standard supergravity (Type I SUGRA)
with the GS correction $\sim \alpha' $ has been realized at the mass shell
in papers \cite{BBLPT}, \cite{AFRR}, \cite{P} (see \cite{MANY} for more
complete list of references). The lagrangian has not been constructed
but it becomes clear, that it contains terms $\sim R^2$ and an infinite
number of terms $\sim {\alpha'}^q K^p,$ $ \ \ q\ge 1, \ p \ge 3$. Several
terms of lowest order were found in \cite{RW}.

The connection between standard and dual supergravity in the superspace
approach was mentioned in \cite{AF} where explicit calculations were not
presented. The iterative scheme for  dual transformation in the
component approach was suggested in \cite{BR}.

Few words about  notations. (See Appendix 1 for details).
This study is made in framework of the superspace approach and
our   notations in general correspond to \cite{BW}
(with the change of overall sign in  metric signature).
The short version of this study was
presented in \cite{STZ2} and
notations here correspond in general to that paper
(small differences are self-evident or explained in the text).

We use the computer program "GRAMA"
\cite{STT} written in MATHEMATICA for analytical calculations in
supergravity.

\section{Gravity Sector}

We start from  the derivation of geometrical equations of motion which are
applied equally for usual and dual formulation of supergravity at least for
the case $\alpha' \neq 0$, $\beta'=0$, which is the main approximation
accepted in the following. We use:

1) Geometrical Bianchi Identities (BI's) for the supertorsion ${T_{BC}}^D$:
\be
\label{ST}
D_{[A}{T_{BC)}}^D + {T_{[AB}}^Q\, {T_{|Q|C)}}^D  -
{{\cal R}_{[ABC)}}^D  = 0.
\ee
Here and in the following ${\cal R}_{\ldots}$ means the supercovariant
curvature (calculated with the torsion full spin-connection).

2) The set of constraints \cite{N}, \cite{T1}:
\be
\label{CNSTR}
{T_{\alpha\beta}}^c = \Gamma_{\alpha\beta}^c \, , \ \ \
{T_{a\beta}}^\gamma ={1\over 72}{({\hat X}\Gamma_a)_\beta}^\gamma
\, ,
\ee
where ${\hat X} \equiv X_{abc}\Gamma^{abc} $. The other nonzero torsion
components are:  $T_{abc} \equiv \eta_{cd}
{T_{ab}}^d$ (here $T_{abc} $  is a completely antisymmetric tensor) and
${T_{ab}}^\gamma. $
Furthemore $X_{abc} = T_{abc}/72 $  as it follows from
(\ref{ST}).

2) Commutation relations for supercovariant derivatives $D_A$:
$$ (D_A\, D_B - (-1)^{ab}D_B\, D_A)\, V_C = $$
\be
\label{CR}
 = -\, {T_{AB}}^Q\, D_Q \,V_C -
{{\cal R}_{ABC}}^D\, V_D -({\cal F}_{AB}\,V_C -
(-1)^{c(a+b)}V_C\,{\cal F}_{AB}),
\ee
where $V_C$ is a vector superfield,
${\cal F}_{AB} $ is  a gauge field which is in the algebra
of internal symmetry group $G$,   the supercurvature  ${\cal R}_{ABCD} $
 differs in sign  in comparison with \cite{T1}.
We introduce  "by hands"  the dilatino superfield $\phi$  and use:

3) The most general representation for spinorial derivative of the dilatino
$\chi$-superfield ($\chi_\alpha \equiv D_\alpha \phi$):
\be
\label{DXI}
D_\alpha \chi_\beta = -{1\over2}\,  \Gamma^f_{\alpha \beta} D_f\phi
+(-{1\over 36} \phi T_{abc} + \alpha' A_{abc})\, \Gamma^{abc}_{\alpha \beta},
\ee
where $A_{abc}$ is   an arbitrary  completely antisymmetric
superfield, which is determined later  in terms
of torsion and curvature, when the axionic superfield Bianchi Identities
will be considered. It will be clear later, that the $A_{abc}$-field
contribution as defined in (\ref{DXI}) is really  proportional to $\alpha'$.

The complete set of additional constraints and equations of motion
for superfields of the  supergravity multiplet were derived from
(\ref{ST})-(\ref{DXI}) in \cite{T1}, see also \cite{N}. They
are presented in  Appendix 2. These equations are transformed into
equations for usual fields, if one calculates spinorial derivatives
of  $A_{abc}$-superfield in terms of torsion and curvature (see below),
and takes  zero superspace components.
(In the following we use the same notations for physical
fields and corresponding superfields in the cases when  it can
not lead to a confusion).

\section{ Gauge Sector }

The derivation of gauge matter fields equations is the   standard
procedure in the superspace approach. (For example see
\cite{BBLPT}, \cite{ADR} and references therein).  We
present here some basic results (see \cite{STZ1}).

The Bianchi Identities for the gauge
superfield ${\cal F}_{AB}$ are:
\be
\label{GBI}
 D_{[A}{\cal F}_{BC)} + {T_{[AB}}^Q\, {\cal F}_{|Q|C)}  \equiv 0,
\ee
where
${\cal F}_{AB}  \equiv {\cal F}^J_{AB}X^J $, where
${(X^J)_i}^j $ are anti-hermitean matrices - generators of $G$.

We use the different notations ${\cal F}_{ab}$ and $F_{ab}$
for the supercovariant and usual tangent-space components.
(The connection see below). To solve (\ref{GBI}) on the mass-shell
 the following  constraint is needed:
\be
\label{GC}
 {\cal F}_{\alpha\beta} = 0
\ee
Then,  one can derive  equation of motion in the form:
\be
\label{LE}
 \Gamma^a\,D_a\,\lambda + {1\over12}\, T_{abc}\Gamma^{abc}\, \lambda
 = 0,
\ee
\be
\label{FE}
 D^a\,{\cal F}_{ab}+
T_{ba}\Gamma^a\,\lambda+2\,\lambda\,\Gamma_b\,\lambda=0.
\ee
where $ \lambda $ is a gaugino superfield:
\be
\label{LD}
{\cal F}_{b\alpha}\equiv (\Gamma_b\,\lambda)_\alpha
\ee
(We do not write spinorial indices explicitely in  cases, where their
position can be reconstructed unambiguously).

The
spinorial derivatives of $\lambda$ and ${\cal F}^{ab} $- superfields also
follow  from (\ref{GBI}), (\ref{GC}) (see Appendix 3).

\section{The 3-form Axionic-Field Sector}

The superspace Bianchi Identities for the axionic field take the form:
$$ D_{[A}H_{BCD)} +{3\over2}\,{T_{[AB}}^Q\, H_{|Q|CD)} =$$
\be
\label{HBI}
 = 3\alpha'\,\left( tr\,[{{\cal R}_{[AB}}{{\cal R}_{CD)}}] -
tr[{\cal F}_{[AB}\,{\cal F}_{CD)}] \right)
\ee

The constraint is:
\be
\label{HCNSTR}
H_{\alpha \beta \gamma} =0
\ee
The 3-form $K$ -field, considered in the Intoduction is connected with the
$H$-superfield by the relation:
\be
\label{KH}
 K_{mnp} =
{E_p}^C{E_n}^B{E_m}^A\,H_{ABC}\vert
\ee
Here $ K_{mnp}\equiv e_p^ce_n^be_m^a\,  K_{abc}. $

The mass-shell solution of (\ref{HBI}) which is compatible with
(\ref{ST}) - (\ref{DXI})  can be obtained using the standard procedure
 \cite{AFRR}, \cite{BBLPT}, \cite{RRZ} (see also \cite{ADR}).
 We find nonzero components of
$H_{ABC}$-superfield in the form:
\be
\label{H1}
 {H_{\alpha\beta a}}=\phi\,(\Gamma_a)_{\alpha\beta}
+\alpha' U_{\alpha \beta a}\, ,
\ee
\be
\label{H2}
{H_{\alpha}}_{bc}=-(\Gamma_{bc}\,\chi)_{\alpha} + \alpha'U_{\alpha bc} ,
\ee
\be
\label{H3}
H_{abc}=
-\phi\,T_{abc}+ \alpha'U_{abc}
\ee
The dilaton field $\phi$ is defined by eq. (\ref{H1}). Then
the Bianchi Identity (\ref{HBI}) defines how it "penetrates"
in all the other   $H$-field components. In particular, $\chi = D\phi$
in (\ref{H2}) as a consequence of Bianchi Identities, etc.

The  $U_{\alpha \beta a}$- and  $U_{\alpha ab}$-superfields
do not contain the contribution from gauge fields.
The $ U_{\alpha \beta a} $
is determined in the following form ($U_a$ denotes the matrix
$U_{\alpha \beta a}$,):
$$ U_a = \left(-{14\over9}T^2_{ab} +
{2\over27}\eta_{ab}\,T^2\right)\Gamma^b + $$
\be
\label{U12}
+\left(-{2\over27}T_{aij}T_{klm}- {1\over9}\eta_{ai}D_jT_{klm}-
{1\over 9}\eta_{ai}{T^2}_{jklm} \right)\,\Gamma^{ijklm}
\ee

The term $\sim \Gamma_a $ in $ U_a $
does not fixed by Bianchi Identities (BI's).
As it follows from (\ref{H1}),
the redefinition of such a term leads to the redefinition of
the dilatonic field. Our choice in (\ref{U12}) is different from that in
\cite{STZ2}, namely: $\phi \, \vert_{ref \ \cite{STZ2}}
 \ra $ $(\phi +({2/27})T^2) $.
That leads to simplification of final results.

Any contribution of the form $\sim \theta_{aijklm}\Gamma^{ijklm}, $ where
$\theta_{aijklm} $ is a comletely  antisymmetric tensor,  does not
 fixed by BI for the  $H_{a \alpha \beta} $. Such a contribution is defined
 unambiguosly by  BI for the $H_{\alpha bc} $.

The $U_{\alpha ab} $-field is equal to ($U_{ab}$
 denotes the  $U_{\alpha ab}$):
$$ U_{ab} = 4\Gamma_{i[a}D^i L_{b]} +{2\over3} L^i T_{abi}+ \left(
{7\over18}{\Gamma^{ijk}}_{[a} L_{b]} -\right.$$
$$\left.-{1\over 3}{\Gamma_{ab}}^{ij}L_k +
{2\over3}{\Gamma^{ij}}_{[a}{T_{b]}}^k+
{2\over9}\Gamma^{ijk}T_{ab} -{2\over3}{\Gamma_{ab}}^iT^{jk} \right) T_{ijk}+$$
\be
\label{U21}
+\left(-{1\over2}\Gamma^{ij}L_{[a}-{4\over3}{\Gamma_{[a}}^iL_j
-{20\over3}\Gamma^i{T^j}_{[a}+{8\over3}\Gamma_{[a}T^{ij} \right)T_{b]ij}
\ee

The  $U_{abc}$-field is equal to:
\be
\label{U}
 U_{abc} = U^{(grav)}_{abc} + U^{(gauge)}_{abc}
\ee
where the gauge-field contribution is:
\be
\label{UG}
 U^{(gauge)}_{abc} = - tr(\lambda\,\Gamma_{abc}\,\lambda)
\ee
but the gravity-field contribution is:
$$U_{abc}^{(grav)} = -2\,D^2T_{abc} -6\,D^i{T^2}_{i[abc]}-
6\,{T^{ij}}_{[a}D_bT_{c]ij} -6\,{{\cal R}^{ij}}_{[ab}T_{c]ij}- $$
$$  -6\,{\cal R}_{i[a}{T_{bc]}}^i +4\, T_{[abc]}^3 -
T_{ij}\Gamma_{abc}T^{ij} -12\,T_{i[a}\Gamma_b{T_{c]}}^i -$$
\be
\label{UGRAV}
-6\,T_{[ab}L_{c]} - L_i\Gamma_{abc}L^i -12\,L_{[a}\Gamma_b L_{c]}
\ee

These superfields were
discussed  earlier in \cite{BBLPT}, \cite{AFRR}, \cite{RRZ} using
another parametrization ( another set of constraints).

The $A$-superfield in (2.4)  is also defined unambiguously from
the (2.2)-component
of the Bianchi Identity (\ref{HBI}) (the $(p,q)$-component of
 a superform contains $p$ bosonic and $q$ fermionic indices). We get:
\be
\label{A}
 A_{abc} = A_{abc}^{(grav)} + A_{abc}^{(gauge)}
\ee
where
\be
\label{AG}
 A_{abc}^{(gauge)}=
  -{1\over 24}\,tr(\lambda\,\Gamma_{abc}\,\lambda)
\ee
and
$$ A_{abc}^{(grav)} = -{1\over18}D^2T_{abc}+{1\over36}D^i{T^2}_{i[abc]}-
{1\over36}{T^{ij}}_{[a}D_bT_{c]ij} -{5\over1944}T^2T_{abc}-$$
$$-{5\over108}{T^2}_{i[a}{T_{bc]}}^i + {5\over 54}T_{[abc]}^3
 - {1\over 3888}{\eps_{abc}}^{ijklmnp}T_{ijk}
\left(D_lT_{mnp}  +{5\over2}{T^2}_{lmnp}\right)  $$
\be
\label{AGR}
 - {1\over24}T_{ij}\Gamma_{abc}T^{ij}
-{1\over48}L_i\Gamma_{abc}L^i +{1\over2}L_{[a}\Gamma_bL_{c]}
\ee
The  $A_{abc}$-superfield is a solution
of eq.'s (A2.9), (A2.10'). That provides a  good check of the result.

Now we are ready to discuss equations (A2.1)-(A2.6) in terms of
fields from the supergravity multiplet.
One must use for this purpose  the expression of $T_{abc}$ in terms
of $H_{abc}$-field. This expression follows from (\ref{H3}) as a
perturbative series in $\alpha'$.  By this way one obtains
equations of motion as a series in $\alpha'$.
All the spinorial derivatives from the $A_{abc}$-field can be calculated
in terms of  fields of supergravity multiplet
in the desired (zero) order in $\alpha'$  using exact relations from
Appendix 3. The
supersymmetry transformations also presented as a series in $\alpha'$ in
this case.  The lowest $\sim
{(\alpha')}^0 $-order corresponds to the  supergravity by Chapline-Manton
\cite{CM}. The next $\sim \alpha'$-order was considered explicitly at the
mass-shell by Pesando \cite{P} using  results from \cite{BBLPT},
\cite{AFRR}.
The lagrangian in the $\alpha'$ order has not been constructed.
 (Unfortunately, because of differences in parametrization
and some differences in the approach we are not able to use
intermediate results from \cite{P}).

Calculations in the highest   orders become tremendously cumbersome.
 But, it seems inconsistent  to consider terms  $\sim (\alpha')^p\ \ $
$p\ge 2 $
 because terms $\sim (\alpha')^2$ of $\sigma$-model loop expansion
 were not taken into account.

We don't consider approximate  equations of motion for
standard supergravity in the $\sim \alpha'$-order, because  this program
will be realized exactly (without any expansion in  $\alpha'$)
 for the dual supergravity  in the next section . Then it will
be a simple algebraic problem to come back
 to  the usual Type I supergravity case,
making the dual transformation (see below).

In spite of a  complicated structure of  $H$-field
equations that follow from (\ref{HBI}), (\ref{HCNSTR}),
 one can make a  useful check  of the procedure.
 Note, that  eq. (A2.5)  must be interpreted as the $H$-field
Bianchi Identity.  So it must coincide with the (4,0)- component
 of (\ref{HBI}).  We have checked that is really the case.

 Namely, one can easily
prove, that the difference between (A2.5) and the (4,0) component of
(\ref{HBI}) is equal to
the (4,0)-component of the superform identity \cite{BBLPT}:

\be
\label{IDENT}
D\,U^{(grav)} + V = tr R^2
\ee
where $U^{(g)}_{(0.3)}=V_{(0.4)}=V_{(1.3)} =0$. The components
$V_{(2,2)}$, $V_{(3,1)}$   can be easily calculated from (\ref{HBI}).
Equation (\ref{IDENT}) is identically  satisfied for (2,2), (1,3), (0,4)
components because it is reduced exactly to
that used for {\it definition} of $A$ and $U^{(grav)}$-superfields.
Then equations corresponding to  (4.0), (3.1)-components  follow identically
by algebraic manipulations from the
equations corresponding to  (2,2), (1,3), (0,4)-
components (cf. \cite{BBLPT}).

One more remark is helpful for the following. All the relations in the
 theory under consideration  are invariant under the
scale transformation \cite{W}, \cite{GN2}:

\be
\label{CONF}
 X_j \rightarrow \mu^{q_j} \, X_j
\ee
where $X_j$ is an arbitrary  field,  $q_j$ is a numerical
factor,   $\mu$ is an
arbitrary common factor.  It is a classical symmetry, because the lagrangian
 is also transformed according to (\ref{CONF}) with $q\neq 0$.

  We present  in the Table 1 the transformation rules for
different fields   (the numerical factors in the table are the
values of $q_j$ for each field):

\bigskip

Table 1
$$
\begin{array}{|c|c||c|c||c|c|} \hline
\phi & -1       & T_{abc}& -{1/2} &T_{ab}^\gamma & -{3/4} \\  \hline
 e_m^a &{1/2} &H_{abc}& -{3/2} &\psi_a^\gamma & - {1/4} \\ \hline
 D_a & -{1/2} & N_{abc} & -{1/2} & \chi    & -{5/4}       \\ \hline
D_\alpha & -{1/4} &A_{abc}& -{3/2}& {R_{ab}}^{cd} & -1 \\ \hline
{\cal F}_{ab} & -1 & \lambda & -{3/4} &  {\cal L} & -2 \\ \hline
\end{array} $$

Now we come to consideration of dual supergravity.

\section{The 7-form Axionic-Field Sector}

One can interpret the same  equations (A2.1)-(A2.6) in terms of the
7-form graviphoton superfield $N_{A_1\ldots A_7}$.
 The Bianchi Identity for such a field takes the form:
\be
\label{NBI}
 D_{[A_1}N_{A_2 \ldots A_8)} + {7\over2}\, {T_{[A_1A_2}}^Q \,
N_{|Q|A_3\ldots A_8)} \equiv 0
\ee
We don't introduce the term $\sim \beta'(D X_7)$ in the r.h.s.
 of eq.(\ref{NBI}) according to the  discussion in Introduction.
(Note, that  such a term breaks
 the scale invariance (\ref{CONF})). It is an additional indication that
contribution  $\sim \beta' $ corresponds to loop corrections in the usual
supergravity).
The following nonzero components provide the mass-shell
solution of (\ref{NBI}) which is
consistent with (\ref{ST})-(\ref{DXI}):
\be
\label{NCNSTR}
 N_{\alpha\beta a_1 \ldots a_5} =
 - (\Gamma_{a_1\ldots a_5})_{\alpha\beta} ,
\ee
\be
\label{N3}
 N_{abc} = T_{abc}\, ,
\ee
where $N_{abc}$ is defined by:
\be
\label{N7}
N_{abc} \equiv {1\over7!}\,{\varepsilon_{abc}}^{a_1 \ldots a_7}\,
 N_{a_1\ldots a_7}
\ee
Note the connection between the supercovarian $N$-field and the $M$-field
discussed in the Introduction:
\be
\label{M7}
 M_{n_1\ldots n_7} = {E_{n_7}}^{A_7}\ldots {E_{n_1}}^{A_1}
N_{A_1\ldots A_7}\vert \,
\ee
Here $ M_{n_1\ldots n_7}\equiv
e_{n_7}^{a_7}\ldots e_{n_1}^{a_1}\, M_{a_1\ldots a_7} $.

{\it It is important}, that the solution (\ref{NCNSTR}), (\ref{N3})
 is valid for {\it any}
$A_{abc}$-field, in particular for that, derived in usual supergravity
 (see eq. (\ref{A})).

Using (\ref{N3}) in the equations of Appendix 2 and defining the $A$-field
according to (\ref{A}), with the substitution (\ref{N3}),
 we get the mass-shell description of dual
supergravity in a closed and relatively simple form (as opposed to
the usual supergravity case!).

Using (\ref{N3}) in the eq. (\ref{H3}) we get the duality relation between
the $H_{abc}$ and $N_{a_1\ldots a_7}$-superfields:
\be
\label{DE}
H_{abc} = -\phi\, N_{abc} + \alpha' \,U_{abc}|_{T_{ijk}\ra N_{ijk}}
\ee
where $U_{abc}$ is defined in (\ref{U})-(\ref{UGRAV}).

Now we come to the detailed study of equations of motion and to the
lagrangian construction in the dual supergravity.

\section{Lagrangian for Gauge Fields}

One must change  variables in eq.'s (\ref{LE}),(\ref{FE}) from
the supercovariant to
usual one's with help of the  relations (A1.1)-(A1.10). In particular,
the relation (A1.10) is important (it follows immediately from
(\ref{N3}) and definition of the ${\tilde M}_{abc}$-field in eq.(A1.5) and
(\ref{M7})). The resulting equations take the lagrangian form and the
corresponding lagrangian is equal to (compare with \cite{GN1}, \cite{BR}):
$$ e^{-1}\,{\cal L}^{(gauge)}={1\over g^2}\, tr\, \left[
{1\over4}\,F_{ba}\,F^{ba}-
{1\over{8 \cdot 6!}}\,{\varepsilon}^{a_1 \ldots a_{10}}\,C_{a_1 \ldots a_6}
\,F_{a_7a_8}\,
F_{a_9a_{10}} -\right.$$
$$-\lambda\,{\hat \nabla}\,\lambda
+{1\over24}\,\lambda\,\hat{{\tilde  M}}\,\lambda
-{1\over2}\,\lambda\,\Gamma^a\,\hat{F}\,\psi_a -$$
\be
\label{LM}
\left.- (\lambda\, \Gamma_b\, \psi_a)\,(\lambda\, \Gamma^a\, \psi^b)
+{1\over2}\,(\lambda\,\Gamma^b\,\psi_b)^2+
{1\over2}\,(\lambda\,\Gamma_a\,\psi_b)^2 \right],
\ee
Here the gauge-field coupling constant $g$  is introduced.
Eq. \ref{LM} disagrees with \cite{GN1} in some terms of fourth
order in fermions.

To find the value of $g^2$ in terms of $\alpha',$
 one must consider the gauge-field contribution
in the supergravity-multiplet equations of motion presented in (A2.1)-(A2.6).
This contribution resulted from the $A_{abc}^{(gauge)}$-superfield (see
 (\ref{AG}) and it's
spinorial derivatives).  Just the same contribution one must  obtain,
 making the variation of ${\cal L}^{(gauge)}$ over the fields of
 gravity multiplet. The comparison of these contributions
  makes it possible to find  $g^2 $ and
to establish  linear combinations of equations (see below relations
(\ref{5.2}) -(\ref{5.4})) that
follow from the lagrangian. It is sufficient to put  $A_{abc}^{(grav)} =0$
 at this stage.
 We get:
\be
\label{ALPHA}
\alpha' = -{1 \over 4\, g^2}
\ee
This relation follows  from the consideration of gauge matter
contribution to the graviton and (independently) gravitino equations of
motion.

\section{Zero Order Lagrangian for Gravity}

It is instructive now, as a first step, to discuss  the gravity-part of a
total  lagrangian  in the limit $\alpha' \ra 0 $ starting from
 equations (A2.1) - (A2.6). It is possible to write this lagrangian in a
simple form  \cite{Z1},  which follows from the linearity
 in $\phi$ and $\chi$ -fields of  the equations from Appendix 2:
\be
\label{LS0}
 e^{-1}\,{\cal L}^{(grav)}_0 = \phi\,({\cal R}-{1\over3}\,T^2)\,
 \vert +2\chi\,\Gamma^{ab}T_{ab}\, \vert \,
\ee
The symbol $\vert$ means as usual the zero superspace-component of
a superfield. The complete explicit result for $ {\cal L}^{(grav)}_0$
as derived  in \cite{Z1} takes the form:

$$e^{-1}\, {\cal L}_0^{(grav)} =
 \phi\,R  -  {1\over 12}\,\phi\,{\tilde M}_{abc}^2 -
 2\,\phi_{;\,a}\psi^a\Gamma_b\psi^b +
4\, \psi_a\Gamma^{ab}\chi_{;\,b} +                                           $$
$$-  2\,\phi\,\psi_a\Gamma^{abc}\psi_{c;\,b}
  + {1\over 12}\,\phi\,\psi_a\Gamma^{[a}
{\hat {\tilde M}}\Gamma^{b]}\psi_b -
 {1\over 2}\,\chi\,\Gamma^{ab}\psi^c {\tilde M}_{abc} -   $$
$$ - {1\over 48}\,\phi\,{(\psi^d\Gamma_{dabcf}\psi^f)}^2 +
{1\over 4}\,\phi\,{(\psi_a\Gamma_b\psi_c)}^2 +
{1\over 2}\,\phi\,(\psi^a\Gamma^b\psi^c)(\psi_a\Gamma_c\psi_b) -           $$
\be
\label{L0}
 - \phi\,{(\psi_a\Gamma^b\psi_b)}^2 +
(\chi\Gamma_{ab}\psi_c)(\psi^a\Gamma^c\psi^b) -
2\,(\chi\Gamma_a\Gamma_b\psi^b)(\psi^a\Gamma_c\psi^c)
\ee
Up to field redefinitions it is the lagrangian obtained in
\cite{BRW}, \cite{GN2}, \cite{GN1}.
Now we are ready to establish  the relation between standard
variables (see Introduction) and that used in the superspace approach.
 In particular:
$\phi = \exp(2\varphi/3) $, $k^2 =1/2 $. The change of variables to the
primed ones, that transforms (\ref{L0}) to the canonical form, is defined by:
$$ e_m^a = \phi^{-1/8}\, {e_m^a}' $$
\be
\label{CHAHGE}
 \chi =
  -{2\over 3 \sqrt 2}\phi^{17/16} \, \chi' \ \ \
  \psi_m = {1\over2}\,\phi^{-1/ 16}\,({\psi_m}'
 - {1\over 6 \sqrt 2}\,{\Gamma_m}'\,\chi') \ \ \ etc.
\ee

The variation of $ {\cal L}^{(gauge)}+ {\cal L}_0^{(grav)} $
  with respect to the
 gravitino field $\psi_m$ produces the equation (see Appendix 2 for
notations):
\be
\label{5.2}
 Q_a + \Gamma_a \, Q =0.
\ee
The variation of the same object   with respect to the axion field
$C_{m_1\ldots m_6}$ produces the equation:
\be
\label{5.3}
 S_{abcd} + 3\, \psi_{[a}\, \Gamma_{bc}\, Q_{d]} = 0.
\ee
The variation   with respect to the graviton field
$e_m^a$ produces the equation:
$$ S_{ab} + \eta_{ab}\, ({1\over 2}B-S)-2\psi_{(a}Q_{b)}-
{1\over 2}\psi^c \Gamma_{ab}\,Q_c - \psi^c \Gamma_{c(a}Q_{b)} -$$
\be
\label{5.4}
 -{1\over 2}(\psi^c \Gamma_{cab} + 2\psi_a\Gamma_b)Q
- \eta_{ab}\,  \psi_c \Gamma^c Q
 -{1\over2}\, \eta_{ab}\, Tr(\lambda \Lambda)-
{1\over4}\, Tr(\lambda \Gamma_{ab} \Lambda) = 0.
\ee
Here $ B\equiv  -\phi ({\cal R} - {1\over 3}\, T^2), \ $ but
  $ \Lambda \equiv ({\hat \nabla} \lambda + \ldots) =0 $ is
the l.h.s. of the gaugino equation (\ref{LE}).
 The  variation  with respect to the dilaton $\phi$
and the dilatino $\chi$-fields produces the constraints (A2.7),
(A2.8).

Calculating $\Gamma_a $ projection from (\ref{5.2}) one immediately
obtains $Q=0$, and then $Q_a=0$. So, $ S_{abcd}=0 $ as it follows from
(\ref{5.3}).  Contracting $a,b$ indices in (\ref{5.4}) one obtains
$S=0$, and then $S_{ab}=0$.
  So, all the equations (A2.1)-(A2.5) follow from (\ref{5.2})-(\ref{5.4}).
 (Equation (A2.6) is equivalent to the Bianchi Identity
 for the $M$-field).

Now we come to  consideration of $\alpha'$ contributions in
pure gravity sector and to the construction of total lagrangian.

\section{Total Lagrangian}

The supersymmetric lagrangian of dual supergravity takes the form
\be
\label{LTOT}
{\cal L} =
{\cal L}^{(gauge)}+ {\cal L}^{(grav)}
\ee
where:
\be
\label{LTOT1}
{\cal L}^{(grav)} =
 {\cal L}_0^{(grav)} + \alpha' {\cal L}_1^{(grav)}
\ee

Now  we are interested in the last term in (\ref{LTOT1}).
It is a property of our parametrization that  ${\cal L}^{(grav)}_1$
  does not depend on $\phi$ and $\chi$ fields. It means that the
  scale invariance
(\ref{CONF})  greatly simplify the possible structure  of this term.

  We consider  the bosonic part  of ${\cal L}^{(grav)}_1$
  which contain 12 possible terms \footnote{K.N.Zyablyuk, unpublished}:

\be
\label{LSUM}
 {\cal L}^{(grav)}_1  = \sum_{i=1}^{12} x_i\, L_i + \mbox{fermions}
\ee
where $x_i$ are numbers to be determined by comparison with equations
(A2.1)-(A2.6), but $ L_i$ are presented in the Table 2.

\centerline{Table 2}
  $$
\begin{array}{||c|c||c|c||c|c||} \hline
i & L_i       & i  & L_i  & i & L_i \\  \hline \hline
 1 & R^2    & 5 & ({\tilde M}^2)\,R & 9 &
  {\tilde M}^{abc;d}({\tilde M}^2)_{abcd} \\ \hline
 2 & R_{ab}^2 & 6 &  ({\tilde M}^2)_{ab}\, R^{ab} & 10
 & ({\tilde M}^2)^2 \\ \hline
3 & R_{abcd}^2 & 7 & ({\tilde M}^2)_{abcd}R^{abcd} & 11  &
  ({\tilde M}^2)_{ab}^2   \\ \hline
4 &\varepsilon^{0\ldots9}R_{01bc}{R_{23}}^{bc} C_{4\ldots9} & 8
 & {\tilde M}^{abc} \nabla_d\nabla^d {\tilde M}^{abc}     &
12 & ({\tilde M}^2)_{abcd}\,({\tilde M}^2)^{acbd} \\ \hline
  \end{array} $$
where $ ({\tilde M}^2)_{ab} = {{\tilde M}_a}^{cd} {\tilde M}_{bcd} $
 and $ ({\tilde M}^2)_{abcd}=
 {{{\tilde M}_{ab}}}^f {\tilde M}_{cdf}$.

Now we come to the determination of $x_i$ in (\ref{LSUM}).
All the terms, containing ${\tilde M}_{abc}$-field  can be
 reconstructed with the help of the following simple procedure.
As was discussed before,  eq.  $S_{abcd}=0$    is equivalent to the
 (4,0)-component of the
$H$-field Bianch Identity, which (dropping spinorial terms) takes the form:

$$ D_{[a}H_{bcd]} +{3\over2}\,{T_{[ab}}^f\, H_{|f|cd]} =$$
\be
\label{HBI40}
 = 3\,\alpha'\,
\left( {{\cal R}_{[ab}}^{ef}{{\cal R}_{cd]}}_{fe}
- tr[{\cal F}_{[ab}\,{\cal F}_{cd]}]\right)
\ee
Changing notations to  the  covariant derivative
  $\nabla_a$, to  the standard  curvature
 $R_{abcd}$ an to the gauge-field $F_{ab}$,
  one can write (\ref{HBI40}) in the form:

$$ \left( H_{[abc} - 3\,T_{ij[a}{R_{bc}}^{ij}
 + {3\over 2} T_{ij[a;b}{T_c}^{ij}
+{1\over 2} T_{ij[a}{(T^2)_{bc}}^{ij}\right){}_{;\,d]} = $$
\be
\label{TEQ}
 = 3\,\alpha'\,\left(-{R_{[ab}}^{ij} R_{cd]ji} +
tr \, [ F_{[ab}\, F_{cd]}] \right)
\ee
Here $;d$ denotes the covariant derivative $\nabla_d $.

Then one can  use (\ref{DE}) and relations from Appendix 1,
 writing everything in terms of
${\tilde M}_{abc}$-field.
 After that the terms in the lagrangian,  containing the
 ${\tilde M}_{abc}$-field,
are immediately reproduced from the l.h.s. of (\ref{TEQ}) which has the
desired form of complete derivative. The terms $\sim MR^2$ and $\sim MF^2$
are  reproduced from the r.h.s. of (\ref{TEQ}).

To check the result and to obtain another terms in the lagrangian, one needs
the explicit form of equations in Appendix 2. So, one must
calculate the first and second  spinorial derivatives from the
$A_{abc} $-field. We have done this calculation. See the result in
Appendix 4. We are  able to present explicitly the  dilaton
equation (in the complete form) and the graviton equation with
bosonic field contributions only.

Such a calculation is possible  only with the help of a computer.
 We use the program "GRAMA"
written by us in "MATHEMATICA" environment which makes it possible
to perform  calculations effectively  using the  PC-486 with RAM=16 Mb.

 One can  obtain
 terms, containing ${\tilde M}$-field  and  other terms $\sim R^2$ in
(\ref{LSUM}) by the following  way. Calculating the variation
of ${\cal L}^{(grav)}$ over the graviton field one  must get the
equation (\ref{5.4}).
Contracting indices $a,b$ one gets the dilaton equation $S=0$, as it is
  explained in Appendix 2.  Comparing
the result with the equation (A4.1) one  finds  values of $ x_i $ in
 eq. (\ref{LSUM}).
The result of this calculation is in  complete correspondence
 with the previous one, based on eq. (\ref{TEQ}).
 The obtained values of $x_j$  are presented in the
Table 3.

\bigskip
Table 3
  $$
\begin{array}{|c|c||c|c|} \hline
x_1 & undetermined   &   x_7 & 0  \\ \hline
 x_2 & 2  & x_8 & -1/6   \\ \hline
x_3 & -1  & x_9  & 1/2   \\ \hline
x_4 & (2\cdot 6!)^{-1}  & x_{10} &  x_1/144 \\ \hline
x_5 & - 2\,x_1/12 &x_{11} & 0  \\ \hline
x_6 & -1/2 & x_{12} & -1/24 \\ \hline
  \end{array} $$
Then we ensure, that graviton  equation (A4.2) follows from
the total lagrangian (\ref{LTOT}). That provides a complete check
of the result.

Terms  containing $x_1$ in (\ref{LSUM}) enter in the combination
$x_1 \, (R-{1\over12}\,M^2)^2 $ which is the square of the constraint
(A2.8) (remind that we are working in the  zero order in fermionic fields,
remind also the difference between  $R$ and ${\cal R}$ fields).
 This constraint is
satisfied automatically on the mass-shell. That is the reason why
$x_1$ is undetermined. One can  put $x_1$ equal
to zero without any effect on  equations of motion.  There is
another argument to omit terms $\sim x_1$: one can
cancel all such terms  by the off-shell $\phi$-field
redefinition: $ \phi \rightarrow  \phi - x_1\cdot\,\alpha' \,(  R -
{1\over 12}{\tilde M}^2) $.

    Finally, (puting $x_1 $ =0)
one can  write the  bosonic part of the gravity lagrangian
(\ref{LTOT1}) in the  form:
$$ {\cal L}^{(grav)} =  \phi\, (R- {1\over 12}{\tilde M}^2) +$$
$$+\alpha'\, \left[ 2\,R^2_{ab} -R_{abcd}^2 +
{1\over 2\cdot 6!}
\varepsilon^{abcdf_1\ldots f_6}\,{R_{ab}}^{ij} R_{cdij}C_{f_1\ldots f_6}
-{1\over 2}\,R^{ab}({\tilde M}^2)_{ab}-\right.$$
\be
\label{L}
\left. -{1\over 6}\, {\tilde M}^{abc}\nabla_f\nabla^f{\tilde M}_{abc} +
 {1\over 2}\, {\tilde M}^{abc;d}({\tilde M}^2)_{abcd}
  -{1\over 24}\, ({\tilde M}^2)_{abcd}({\tilde M}^2)^{acbd}\right] +
\mbox{spinors}
\ee

One can restore the ghost-free Gauss-Bonnet combination \cite{Z}:
$R^2_{abcd}-4\,R_{ab}^2+R^2 $, adding the square of the constraint
(A2.8) and the square of the graviton equation (A2.4) to the lagrangian
(\ref{L}). It does not change equations of motion but makes the
lagrangian much more cumbersome. It is one of the reasons why attempts
to supersymmetrize the Gauss-Bonnet combination started in \cite{RW}
were not succesful.

One can make by a standard procedure the dual transformation in the total
lagrangian (\ref{LTOT}), adding the term with the lagrangian multiplier
$B_{mn}$:
\be
\label{LMULT}
\Delta {\cal L} = {1\over2}B_{mn}\partial_n {\tilde M}^{mnp}=
-{1\over6}\left[K_{mnp}-2\,\alpha'(\Omega_G-\Omega_L)_{mnp}
\right]\,{\tilde M}^{mnp}
\ee
After that, one can consider ${\tilde M}_{abc}$ as an independent variable.
Solving the  equation of motion for ${\tilde M}_{abc}$, one is able to
reproduce the bosonic part of dual transformation (\ref{DE}) with
$ K_{abc}-2\, \alpha' \,X_{abc} $  instead of $H_{abc}$ in the
l.h.s. of (\ref{DE}). Here $X_{mnp}$ is a 3-form field defined in (A1.13).
The term $2\alpha'X_{abc}$ describes the  change of the CS-term in $K_{mnp}$
to that defined by torsion-full spin-connection.  So, there is a
complete correspondence between (\ref{LTOT}) and dual transformation
(\ref{DE}). Using (\ref{DE}) one can rewrite the lagrangian in terms
of $K_{abc} $-field, obtaining the lagrangian for standard
formulation of supergravity. But in this case the result can be
 presented as an infinite series in $(\alpha')^p, \ $  $p\ge 1$.

\bigskip

{\bf \Large Appendix 1}

\bs

{\bf Notations}

\bs

The 10-dimensional metric signature is:  $\eta_{ab} =diag(1,-1, \ldots -1)$.
The tangent-space vector indices are from the beginning of the
alphabet:  $a,b, \ldots $; the world indices are from the middle of the
alphabet: $ m,n, \ldots $. We use the 16-components spinors and
spinorial indices are $\alpha, \beta, \ldots $. Gamma-matrices are
$\Gamma^a_{\alpha \beta}, ({\Gamma^{ab})_\alpha}^\beta $, etc.
The algebraic properties  of $\Gamma$-matrices are described in many papers.
We use extensively  relations from \cite{BBLPT}, \cite{T2}.
The superspace indices are $ M=(m, \mu),\  N=(n, \nu), \ldots $ and
$A=(a, \alpha),\  B=(b, \beta), \ldots $.

We use  standard conventions: $ tr\,F\wedge F =d\Omega_{G} $,
  $\  tr\, R\wedge R =d\Omega_{L}  $,
 where  symbol $tr$ means $trace $ in the vectorial representation
of the corresponding group, i.e. the  $O(1,9)$-group for $\Omega_L$ and
$SO(32)$ for $\Omega_G$. One must change $tr \ra (1/30)Tr $
 for the gauge-group  $E_8\times E_8 $, where  $Tr$ means  $trace$ in the
adjoint representation. The same change is possible also for $SO(32)$.
Usually  we drop the $\wedge$-sign in products of forms.

The Lorentz Chern-Symons term is defined by:
 $\Omega_L =tr\,(\omega d\omega + 2\,\omega^3/3)$, where $\omega$ is the
spin-connection 1-form, $R=d\omega +\omega^2 $. The same expressions
are used for the gauge-group  CS-term $\Omega_{G}$ and the gauge-field $F$
in terms of the 1-form potential $A$.

The following  notations are used in (A2.1)-(A2.10) below and in the
main text:
$$  L_a = T_{ab}\Gamma^b,  \ \ {\hat Z} = Z_{ijk}\Gamma^{ijk} $$
$$  YZ = Y_{ijk}Z^{ijk}, \ \  (YZ)_{ab} = Y_{aij}{Z_b}^{ij}, \ \
   (YZ)_{abcd} = Y_{abj}
{Z_{cd}}^j,    $$
$$ Z^3_{abc} = Z_{aij}{Z_b}^{jk}{Z_{ck}}^i $$
where $Y, \, Z$ are 3-rd rank antisymmetric   tensors.

We present also  relations between supercovariant and space-time
covariant objects. The gravitino field
$\psi_b^\alpha $ is defined with the help of the superspace veilbein
(cf. \cite{BW} ):
$$ {E_M}^A \vert =
\left(
\begin{array}{ll}

 {e_m}^a & \psi_m^\alpha \\
 0     & \delta_\mu^\alpha

\end{array} \right)\, ,                                   \eqno(A1.1)    $$
The supercovariant derivative $D_a \equiv {E_a}^M\,D_M $ is equal to:
$$D_a = e_a^m\,D_m -\psi_a^\beta\,D_\beta \, ,              \eqno(A1.2)$$
where $\psi_a = e_a^m\, \psi_m\, , $
the space-time component of the covariant
derivative is:
$$ D_m \lambda = \partial\, \lambda - \phi_m \, \lambda -
[A_m, \lambda],                                              \eqno(A1.3)  $$
where ${(\phi_m )^\beta}_\gamma \equiv {1\over 4}
{\phi_m}^{ab}{({\Gamma_{ab}})^\beta}_\gamma   $ is the
spin-connection which is in the algebra of $O(1.9)$.

We introduce also the usual tangent-space components of physical
fields  instead of supercovariant quantities.
(Note, that supercovariant components are equal to:
${\cal F}_{ab}$  $= E_a^ME_b^N \,{\cal F}_{MN},$ etc.).
  Namely:
$$F_{ab} \equiv e_a^m\,e_b^n\,F_{mn}, \ \ \ \
\omega_{cab} \equiv e_c^m\, \omega_{mab}\, ,               \eqno(A1.4) $$
$${\tilde M}_{abc}=
{1\over7!}\,{\varepsilon_{abc}}^{a_1 \ldots a_7}\,({e_{a_1}}^{m_1}
\,\ldots {e_{a_7}}^{m_7}\, M_{m_1\ldots m_7})\, ,               \eqno(A1.5) $$
where $ M_{m_1\ldots m_7}= 7\, \partial_{[m_1}\,C_{m_2 \ldots m_7]}$, and
 $C_{m_1\ldots m_6}$ is the 6-form axionic potential.

 One   finds the relation by a standard procedure
 between the torsion-full
spin-connection in  eq.(A1.3) and the usual spin-connection
$\omega_{cab}$ defined in terms of $e_m^a$:

$$ \phi_{cab}= \omega_{cab}(e)   +{1\over2}\,T_{cab}+S_{cab}\, ,
                                                       \eqno(A1.6)$$

 where:

  $$ S_{cab}= \psi_a\,\Gamma_c\,\psi_b - {3\over 2}
\psi_{[a}\,\Gamma_c\,\psi_{b]}                           \eqno(A1.7)$$

 We use also the  notation $  \nabla_m   $  for the covariant
derivative with the spin-connection $\omega_m^{(0)}$
($ \nabla^{}_{[m}e_{n]}^a =0$),   $\nabla_a \equiv e_a^m
 \nabla_m $.

To be complete  we present  the  connection between  physical fields
introduced before  and supercovariant fields (on the mass shell):

$${\cal F}^{ab}=F^{ab}+2\,\psi^{[a}\,\Gamma^{b]}\,\lambda   \eqno(A1.8)$$

$$T_{ab} = 2\,\nabla_{[a}\,\psi_{b]}+
{1\over2}\,(\Gamma^{cd})\,\psi_{[a}\,C_{b]cd} -$$
$$-{1\over72}T_{cde}(\Gamma_{[a}\Gamma^{cde} +3 \Gamma^{cde}\Gamma_{[a})
\psi_{b]}               \eqno(A1.9)$$

$$T_{abc}=
{\tilde M}_{abc}-{1\over2}\,\psi_f\,{{\Gamma^f}_{abc}}^d\,\psi_d
                                                            \eqno(A1.10)$$

$$ {\cal R} -{1\over3}T^2_{abc} = R -{1\over 12}({\tilde M}_{abc})^2 +
\mbox{spinorial \ \ terms}.                              \eqno(A1.11) $$

$$ {\cal R}_{mnab} = R_{mnab} +\nabla_{[m} \,T_{n]ab}-
{1\over2}T^2_{a[mn]b} + \mbox{spinorial terms}          \eqno (A1.12)      $$

In deriving of  (A1.10) relations (\ref{N3}), (A1.5) were used.
${\cal R} $ is defined in terms of spin-connection $\phi$, but
$R$ -in terms of $\omega(e)$

It is instructive to present the relation between the CS-term
${\tilde \Omega} =tr\, (\phi d\phi + 2\phi^3/3)$ and usual
CS-term, defined in terms of $\omega_{mab}(e) $. One gets (in form
notations):

$$ {\tilde \Omega} = \Omega +X +\mbox{spinors}, \ \
 X={1\over12}T^3 +TR +{1\over4}TdT+{1\over4}d(\omega T)  \eqno(A1.13) $$

where one-form field ${(T_m)_a}^b \equiv T_{mac}\eta^{cb}  $ is introduced.

{\bf \Large  Appendix 2}

\bs

{\bf Constraints and equations of motion}

\bs

We present here equations of motion and constraints that follow from
the mass-shell solution of Bianchi Identities.

Gravitino equation of motion:
     $$ Q_a \equiv \phi L_a - D_a\chi - {1\over 36} \Gamma_a {\hat
T}\chi - {1\over 24} {\hat T}\Gamma_a\chi + \alpha'\left(
{1\over 42} \Gamma_a
\Gamma^{ijk}DA_{ijk} + {1\over 7} \Gamma^{ijk}\Gamma_a DA_{ijk} \right)= 0,
                                                            \eqno(A2.1) $$
Dilatino equation of motion:
$$ Q \equiv {\hat D}\chi + {1\over 9}{\hat T}\chi +
 {\alpha'\over 3}\Gamma^{ijk}DA_{ijk} = 0.
                                                              \eqno(A2.2)  $$
Dilaton equation of motion:
$$ S \equiv D_a^2 \phi + {1\over 18}\phi T^2 -\alpha' \left( 2\, TA +
  {1\over 24} D\Gamma^{ijk}DA_{ijk}\right) = 0.
                                                             \eqno(A2.3)  $$
Graviton equation of motion:
 $$ S_{ab} \equiv \phi {\cal R}_{ab} -  L_{(a}\Gamma_{b)}\chi -
 {1\over 36}\phi\eta_{ab}T^2 +
 D_{(a}D_{b)}\phi +$$
 $$+\alpha'\left(-2\, T_{(a}A_{b)} + {3\over 28}D{\Gamma^{ij}}_{(a}DA_{b)ij} -
 {5\over 336}\eta_{ab} D\Gamma^{ijk}DA_{ijk}\right)=0.
                                                            \eqno(A2.4)$$
Axionic equations of motion and Bianchi Identities:
$$S_{abcd} \equiv D_{[a}(\phi T_{bcd]}) + {3\over 2} T_{[ab}\Gamma_{cd]}\chi
 + {3\over 2}
\phi T^2_{[abcd]} +$$
 $$+\alpha'\left( {1\over 12} (T\epsilon A)_{abcd} + 6\,  (TA)_{[abcd]}
 + {3\over 4}D{\Gamma_{[ab}}^jDA_{cd]j}\right) = 0.
                                                           \eqno(A2.5) $$
$$ D^aT_{abc} = 0,                                        \eqno(A2.6)  $$
  There are constraints:
$$T_{ab}\Gamma^{ab} =0,                                   \eqno(A2.7) $$
 $$  {\cal R} - {1\over 3}\, T^2 =0,                      \eqno(A2.8)  $$
where ${\cal R}$ is a supercurvature scalar
 (${\cal R} \equiv {\cal R}_{abcd}\eta^{ac}\eta^{bd}$, \ \
 \   $T^2 \equiv T_{abc}T^{abc}$).
(There are  additional relations, which are not interesting for
our purposes here, see \cite{T1} for details).

Note, that
$$ \Gamma^a Q_a = -Q\, , \ \ \ \eta^{ab}S_{ab} = S   $$
The components of the supercurvature  are defined from (\ref{ST}),
(\ref{CNSTR}) in the form:
$$ {\cal R}_{\alpha\beta ab} =  {5\over6}\, T_{abc}\Gamma^c_{\alpha\beta} +
{1\over 36}\, T_{ijk}\, ({\Gamma^{ijk}}_{ab})_{\alpha\beta} ,
                                                            $$
 $$ {\cal R}_{abc} =  2\, T_{a[b}\Gamma_{c]}
 - {3\over 2} L_{[a}\Gamma_{bc]}   $$
 Furthemore:
  $$  {\cal R}_{[abc]d} =  D_{[a}T_{bc]d} +
   T^2_{[abc]d} , \ \ \ {\cal R}_{[ab]} = 0 ,         $$
There are  two equations for the superfield $A_{abc}$. The
first one follows from the self-consistency of equations of motion (cf.
\cite{N}, \cite{T1}):
$$ {D\Gamma_{[a}}^{ij}D\, A_{b]ij} + 56\, D^jA_{jab}
- {64\over 3} (TA)_{[ab]} =0.
                                                         \eqno(A2.9) $$
The second one \cite{T1}, \cite{BBLPT}  means, that 1200 IR
 contribution to the $A$-field
spinorial derivative is equal to zero:
   $$ (D_\alpha A_{abc})^{(1200)} = 0,
                                                       \eqno(A2.10) $$
It follows from (A2.10):
$$ DA = {\Gamma_{abc}}^{ij}X_{ij}                     \eqno(A2.10')$$
where $X_{ij}^\alpha $ is an arbitrary function which is 16+144+560
 representation of $O(1.9)$. It follows also from (A2.10):
   $$ D_{[a}\,
A_{bcd]} + 2\, (TA)_{[abcd]} +{1\over 360}(T\epsilon A)_{abcd}- $$ $$ -
 {1\over 16\cdot 60}D{\Gamma_{abcd}}^{ijk}D\, A_{ijk} + {1\over 16}
 D{\Gamma_{[ab}}^iD\, A_{cd]i} = 0,
                                                             \eqno(A2.10'') $$

\bigskip

{\bf \Large Appendix 3}

\bigskip
{\bf Spinorial derivatives}

\bs
We present here spinorial derivatives of fields entering
into the $A_{abc} $-superfield in (\ref{AG}), (\ref{AGR}).
 All the
results, written below, follow from  (\ref{ST}), (\ref{CNSTR}),
and (for gauge fields)- from  (\ref{GBI}), (\ref{GC}).
The results for curvature tensor spinorial derivatives were checked
 by us independently  using the curvature tensor Bianchi Identity:
$$ D_{[A}\,{{\cal R}_{BC)D}}^{E} +
 {T_{[AB}}^Q\, {{\cal R}_{|Q|C)D}}^{E} =0         \eqno(A3.1) $$
Spinorial derivative of the curvature superfield is:
$$ D {\cal R}_{abij} = 2\, D_{[a}\, {\cal R}_{b]ij} +
{1\over36}\, T^{mns}\, \Gamma_{mns}\Gamma_{[a}\,
 {\cal R}_{b]ij} +$$
$$+ {\cal R}_{dij}\, {T_{ab}}^d
- \left({5\over6}\, T_{ijk}\Gamma^k +
{1\over36}\, T_{mnp}{\Gamma^{mnp}}_{ij}\right)\,
T_{ab}                                                \eqno(A3.2)$$
where:
$$ {\cal R}_{abc} = 2\,\Gamma_{[b}\, T_{c]a} + {3\over 2}
  \Gamma_{[ab}\,L_{c]}                                        \eqno(A3.3) $$

Spinorial derivatives of the torsion superfield are:
$$ D\,T_{abc} = 3\,\Gamma_{[a} T_{bc]}+
 3\,\Gamma_{[ab} \,  T_{c]}                                   \eqno(A3.4) $$
$$ D_{\alpha}\,(T_{ab})^{\beta} =
 {({\hat O}_{ab})^\beta}_\alpha                            \eqno(A3.5) $$
where:
$${\hat O}_{ab} =-{1\over 36}\Gamma_{[a} \Gamma^{ijk}D_{b]}\, T_{ijk}
+ {1 \over 36 \cdot 72}\,
 \Gamma_{[a} \Gamma^{mnp}\Gamma_{b]}  \Gamma^{ijk}\,
T_{mnp}T_{ijk} -$$
 $$+ {1\over 72}\, \Gamma^m  \Gamma^{ijk}  T_{abm}T_{ijk}-
{1\over 4} {\cal R}_{abij}\Gamma^{ij}                        \eqno(A3.6)   $$

Spinorial derivatives of matter fields are:
$$ D_\alpha\lambda^\beta =
  {1\over 4}{\cal F}_{ab}{(\Gamma^{ab})_\alpha}^\beta  \eqno(A3.7)$$
and:
$$ D\,{\cal F}_{ab}= 2\Gamma_{[a}\,
D_{b]}\,\lambda-
T_{abc}\,\Gamma^c\,\lambda
-{1\over36}\,
 T^{ijk}\Gamma_{ijk}\Gamma_{ab}\,\lambda                   \eqno(A3.8)$$

\bigskip

{\bf \Large Appendix 4}

\bigskip

{\bf Dilaton and graviton equations of motion}

\bs

We present here in explicit form  the dilaton  equation and bosonic
contribution to the graviton  equation of motion as they follow from
(A2.3) and (A2.4). The dilaton equation:
$$D^2\phi +{1\over18}\phi\,T^2 -{1\over 3}\alpha' \, tr\,
{\cal F}_{ab}^2  -\alpha'
 \left[
  -{2\over3}({\cal R}_{ab})^2
+{1\over3}({\cal R}_{abcd})^2 +{1\over3}{\cal R}^{ab}T^2_{ab}-
 \right. $$
$$-{1\over3}
{\cal R}^{abcd}T^2_{abcd}+{1\over9}T^{abc}D^2T_{abc}-{1\over6}D^2T^2+
{1\over3}D^aD^bT^2_{ab}+ $$
$$+{4\over3}L^a{\hat D}L_a +{8\over3}T^{ab}D_aL_b +
T^{abc}
 \left(
 -{4\over3}L_a\Gamma_bL_c +{5\over54}L^d\Gamma_{abc}L_d-
\right. $$
$$\left. \left.
-{1\over9}L^d\Gamma_{ab}T_{cd}-{1\over54}T^{df}\Gamma_{abc}T_{df}+
{7\over3}T_{ab}L_c +{2\over3}{T^d}_a\Gamma_bT_{cd}
 \right) \right] =0
                                                              \eqno(A4.1)$$
The graviton equation:
$$    \phi{\cal R}_{ab} +D_{(a}D_{b)}\phi - {1\over36}\eta_{ab}\,\phi T^2
-\alpha' \, tr\,\left( 2\, {{\cal F}_a}^j {\cal F}_{bj} - {1\over6}
\eta_{ab} {\cal F}_{ij} {\cal F}^{ij}\right) - $$
$$-\alpha'\Biggl[ -4\,{{\cal R}_a}^c{\cal R}_{bc} +
2\,{\cal R}_{ijka}{{\cal R}^{ijk}}_b  -  2\,D^2{\cal R}_{ab}- $$
$$-{1\over 6}D_aT_{ijk}D_bT^{ijk}
+D_iT_{jk(a} \left(D_{b)}T^{ijk}-{1\over2}D^i{T_{b)}}^{jk} -
D^k{T_{b)}}^{ij} \right) + $$
$$     + {T^2}_{ijk(a}\left(2\,D_{b)}T^{ijk} - 4\,D^i{T_{b)}}^{jk}-
2\,D^k {T_{b)}}^{ij}\right) +4\, {{T_a}}^{ij} {T_b}^{kl} {T^2}_{ikjl} -$$
$$   -2\,{{{{{T^2}_a}^i}_b}^j}{T^2}_{ij} +\eta_{ab} \left(
{1\over3}({\cal R}_{ij})^2 -{1\over6}({\cal R}_{ijkl})^2 -
{1\over6}{\cal R}^{ij}{T^2}_{ij} +
{1\over6}{\cal R}^{ijkl}{T^2}_{ijkl}-\right. $$
$$
\left. -{1\over18}T^{ijk}D^2T_{ijk} -{1\over6}D^iD^j{T^2}_{ij} +
{1\over12}D^2T^2 \right) \Biggr] + \mbox{fermions} = 0
                                                                 \eqno(A4.2)$$

{\bf \Large Appendix 5}

\bs

{\bf Supersymmetry transformations}

\bs

Supersymmetry transformations for any physical field follow
 immediately  from the super-gauge transformation for the
corresponding superfield \cite{BW}. We get for gauge matter
multiplet:

 $$ \delta_Q(\epsilon) \, \lambda =
  {1\over4}\, {\cal F}_{ab} \, \Gamma^{ab}\, \epsilon    $$
 $$ \delta_Q(\epsilon) \, A_m = -\lambda \, \Gamma_m \, \epsilon,
  \eqno(A5.1) $$
where $\eps^\alpha $ is a parameter,
 $\Gamma_m \equiv e_m^a \,  \Gamma_a \ $.

For  gravity multiplet :

$$\delta_Q(\epsilon) {e_m}^a =- \psi_m\Gamma^a\epsilon\, ,    $$
$$ \delta_Q(\epsilon) \psi_m =- D_m \epsilon - {1\over 72}\,
 \Gamma_m{\hat T}\,\epsilon\, ,                         $$
$$ \delta_Q(\epsilon) \phi =  \chi\,\epsilon\, ,                $$
$$ \delta_Q(\epsilon) \chi =  {1\over 2}\, \partial_m \phi \,\Gamma^m
-{1\over2}(\psi_m \, \chi)  \,(\Gamma^m\epsilon) -
({1\over 36}\,\phi{\hat T} -\alpha'
{\hat A} )\, \epsilon\, ,                    $$
$$ \delta_Q(\epsilon) C_{m_1 \ldots m_6} =  6\,\psi_{[m_1}
\Gamma_{m_2 \ldots m_6]}\,\epsilon\,                 \eqno(A5.2) $$

  It is the advantage of our
parametrization, that matter degrees of freedom as well as
superstring $\sim \alpha'$ corrections "penetrate" the gravity multiplet
supersymmetry transformations only due to the
$A_{abc}$-contribution.

The supersymmetry algebra for physical fields
is closed up to  equations of motion
and  gauge transformations. Namely:

$$ [\delta_Q(\epsilon_2),\, \delta_Q(\epsilon_1)]\,X =\left(
 \delta_{GCT}(\xi^m)+  \delta_Q(\epsilon')   + \delta_L(L_{ab})+
\right. $$
$$  \left.
+ {\delta_G}(\Omega_{YM}) + {\delta_A}(f_{n_1\ldots n_5})
  \right)\, X +
 (\mbox{equations of motion}),   \eqno(A5.3)        $$
where $X$ is any field from gravity or  matter multiplet,
$\delta_{GCT} $ is a general coordinate transformation,
$\delta_L$ is a Lorentz transformation, $\delta_G$ is a matter
gauge-field transformation, $\delta_A$ is an axion-field
gauge transformation.

The  transformation parameters  are:

$$ \xi^m = \epsilon_1\, \Gamma^m \, \epsilon_2\, .       $$
$$ \Omega_{YM} = -  \xi^m \, A_m\, .     $$
$$ \Omega_{m_1,\ldots, m_5} = -\xi^n\,C_{m_1,\ldots, m_5,n}  $$
$$ L_{ab} = - \xi^n \, \phi_{nab} + {5\over 12}\, \xi^c\, T_{abc} +
{1\over 36}\, \epsilon_1\, {\Gamma_{ab}}^{ijk}\, \epsilon_2\, T_{ijk}.
  $$
$$ \epsilon' = \xi^n\, \psi_n                     $$
 Eq. (A5.3) takes place for any $A_{abc}$-field, not
specifically for that, defined by eq.(\ref{A}).  Only the
representation (A2.10') for the $A_{abc}$-superfield spinorial
derivative is necessary  for the derivation of (A5.3) .

\end{document}